# Bound exciton and free exciton states in GaSe thin slab

Chengrong Wei[1†], Xi Chen[2†], Dian Li[2], Huimin Su[1], Jun-Feng Dai[1]*

1 Department of Physics, South University of Science and Technology of China, Shenzhen 518055, China

2 Physics Department, The University of Hong Kong, Pokfulam road, Hong Kong, China

† These authors contributed equally to this work.

* daijf@sustc.edu.cn

Abstract: The photoluminescence (PL) and absorption experiments have been performed in GaSe slab with incident light polarized perpendicular to c-axis of sample at 10K. An obvious energy difference of about 34meV between exciton absorption peak and PL peak (the highest energy peak) is observed. By studying the temperature dependence of PL spectra, we attribute it to energy difference between free exciton and bound exciton states, where main exciton absorption peak comes from free exciton absorption, and PL peak are attributed to recombination of bound exciton at 10K. This strong bound exciton effect is stable up to 50K. Moreover, the temperature dependence of integrated PL intensity and PL lifetime reveals that a non-radiative process, with active energy extracted as 0.5meV, dominates PL emission.

The study of two-dimensional crystals has become a very popular topic in the field of condensed matter physics because of its novel phenomena and the potential for optoelectronic and electronic applications. Recently, a new class of 2D crystals, atomically thin metal monochalcogenides (such as GaSe), has attracted much attention. This kind of material exhibits various interesting physics such as magnetism in p-type monolayer GaSe [1], non-Markovian memory effects [2], strong SHE effect in multilayer GaSe [3-5], optical spin polarization[6, 7] and novel behavior of Landau levels in the presence of a Coulombic electrons-hole interaction of bulk GaSe [8-12], . Actually, the study on GaSe 2D crystal could be traced back to 1958, in which the exciton spectrum near fundamental absorption edge of bulk GaSe was first reported [13]. Free and bound exciton behavior in this system has always been discussed [14, 15] since its discovery. Here we report the remarkable energy difference between absorption peak and photolumininescence (PL) peak of GaSe slab at low temperature, which is about 34meV at 10K. By studying the temperature dependence of PL spectra, we attribute the absorption peak to free exciton absorption, and the PL peak to bound exciton recombination. So the energy difference of 34meV origins from energy difference between free exciton and bound exciton states. Meanwhile, the temperature dependence of PL integral intensity and PL lifetime shows that a non-radiative process contributes most in the PL emission at the temperature above 50K.

The crystal of GaSe consists of covalently bonded layers by van der Waals stacking, and each layer contains four monoatomic sheets in the sequence of Se-Ga-Ga-Se. Bulk GaSe is generally known as an indirect semiconductor, with the lowest conduction band minimum at M point and the highest valance band maximum at $\Gamma$ point of the Brillouin zone. However, the energy difference between $\Gamma$ and M points at the lowest conduction band is in the range of 10-20meV and the direct gap is just slight higher than the indirect gap. Due to weak interlayer interactions and high electron density within layers, exciton effect is prominent in monolayer and bulk GaSe material. According to Mooser`s theory published in 1973, optical selection rules near $\Gamma$ point could be explained by using an exciton (two-particle) picture. When spin-orbit coupling is ignored, orbital symmetries only allow direct transition near $\Gamma$ point for the case of E//c, where E is the polarization of light and c is the c-axis of crystal. When spin-orbit interaction is considered, electron transition between the s-like upper most valence band and p-like lowest conduction band near the $\Gamma$ point is also dipole-allowed for the geometry of $E \perp c$. However, the possibility of it is about 30 times smaller than that in the case of the excited light with $E // c$. All of our experimental measurements are based on the geometry of $E \perp c$.

The photoluminescence and absorption spectra were collected in transmission geometry with temperature ranging from 7K to room temperature. During the PL measurement, GaSe slab was normally excited by a CW laser at wavelength of 532nm. While in the absorption measurement, a supercontimuum white light laser was used. The PL and absorption spectra were obtained by a spectrometer equipped with a cooled charge-coupled device (CCD). The direction of electric field of excited light for both PL and absorption measurement was perpendicular to the c axis of samples. Besides, ultrafast PL spectra were also employed to study dynamics of electron-hole pairs of GaSe material. Samples were excited by a pulsed laser with wavelength of 560nm, pulse widths of 150fs and pulse repetition rates of 80MHz. The time-resolved PL spectra were measured by a single photon avalanche diode with photon timing resolution of 50ps.

The black curve in Fig. 1a shows the absorption spectrum of a GaSe slab at 10K. An unambiguous peak (2.12eV) near the fundamental absorption edge of GaSe is observed under T=10K, which can survive even at room temperature. We attribute this sharp absorption peak to an excitonic transition, which was firstly observed in bulk GaSe in 1958 by Fieldingl Fischer and Mooser [16]. With increase in temperature, the peak position of exciton emission at band edge shifts from 2.12eV at 10K to 2.01eV at 290K. The temperature dependent energy shift of exciton peak obtained from absorption spectra is shown in Fig. 1b, where black squares are the experimental data and the red line is the theoretical fit. The fitting procedure was performed using Varshni empirical relationship $E_g(T) = E_g(0) - \frac{AT^2}{T+B}$, where $E_g(0)$ is the band gap of GaSe slab at 0K, A and B are constants referred to as Varshni coefficients. The constant A is related to the electron/exciton-phonon interaction and B is related with the Debye temperature of material, and is 251K for GaSe.

PL spectrum (red curve) from GaSe slab at the same temperature (T=10K) is also shown in Fig. 1a. It is observed that the PL spectrum of thin slab consists of multiple lines, where three peaks of PL spectrum are identified at 2.086eV, 2.064eV and 2.032eV respectively (shown in Fig.1c). In PL spectrum, the highest peak energy is at 2.086eV with a FWHM of 14meV, which is 34meV lower than that of exciton peak at 2.12 eV with a FWHM of 10meV in absorption spectrum. There is no observable exciton emission in PL spectrum detected around 2.12eV, which corresponds to the exciton absorption peak in absorption spectrum. In order to rule out the possibility of sample heating by the high-power pump laser applied, which may lead to red shift of PL peak, PL spectrum was also measured under different excitation light intensities. The position of the highest energy peak keeps at around 2.086eV, independent of excitation light intensity ranging from 30uw to 600uw as shown in Fig.1d. The excessive heat induced by absorption could be completely consumed by cold finger.

This discrepancy between the lowest peak position of PL spectrum and that of absorption spectrum is unusual in intrinsic semiconductor, but it has been frequently observed in the heavily doped semiconductors, where the Fermi level lies within the conduction band and the lowest states in the conduction band are filled. Photons energy larger than the fundamental band gap is needed to excite a transition from valence band to unoccupied states in the conduction band in absorption process. While in photoluminescence process, electron-hole recombination occurs near the band edge. Hence, the absorption peak shifts to higher energy than that of photoluminescence peak. If this mechanism works, we may deduce the electron density using equation

$N = \frac{[2m^*(E_F - E_c)]^{3/2}}{6\hbar^3\pi^2}$, where $E_F - E_c$ is the energy difference between Fermi surface and the bottom of conduction band, and $m^*$ is the effective mass near $\Gamma$ point. Ignoring the spin degeneracy, which would result in higher electron density, the electron density N is estimated to be $1.96 \times 10^{18} cm^{-3}$, 2 orders of magnitude larger than that of intrinsic material. However, the GaSe sample measured in our experiments was grown by CVD method without intertional doping. Electric-conductance measurements also show that GaSe slab is electrically intrinsic. Thus, we may safely rule out this possibility.

This phenomenon was also observed in indirect band gap semiconductor. In this case, most absorption occurs in direct transition of electrons around $\Gamma$-point under low temperature. After excitation, electrons (holes) are quickly scattered to the lowest point of conduction band near M-point, and then recombine with holes while releasing phonons. This process could result in the red shift of PL spectrum compared with absorption one. As to an indirect gap material, band-gap photoluminescence is a phonon-assisted process and the quantum yield (QY) is usually negligible comparing to that at direct gap. In order to get PL QY of GaSe slab, a thin film of fluorescent dye was used as PL standard, which was estimated to be 0.511 by using absolute PL quantum yields measurement system. Nevertheless, the PL QY of GaSe slab at 10K is estimated to be about 0.05, which is one order of magnitude larger than that of monolayer MoS2, a direct bandgap semiconductor [17]. From the temperature-dependent PL spectrum as shown in Fig. 3a, PL intensity at 10K is 4 orders of magnitude larger than the one at room temperature. These evidences tend to support the possibility that the PL of our GaSe sample at low temperature is attributed to the direct interband transition. Nonetheless, the experimental result reported by Mosser's group [13] showed that there exists an indirect gap, several tens of meV below the direct one. And the indirect gap is only visible in thick sample with light incident along the layer plane. This is completely different from our experimental setup. So this hypothesis is also excluded.

One potential reason may arise from high density of state (DOS) below the top of valance band, which exhibits a sharp van Hove singularity [1]. The DOS of single layer GaSe is almost a step function at the VBM and gives a quickly peak at the energy 0.013 eV below the VBM. Therefore, the energy of strong exciton absorption peak, which emerges at the position of highest DOS, is higher than that of PL peak resulted from the electron-hole pair recombination near the band gap. However, this unusual character of large DOS at VBM mainly stem from a Mexican-hat-like energy surface around the zone center ($\Gamma$) at the monolayer limit. And according to theoretical calculation, GaSe monolayer has an indirect band gap [18]. It is inconsistent with our experimental result that bulk GaSe is a direct band gap semiconductor at low temperature. So this explanation is also excluded.

Fig. 2a shows the evolution of PL spectra with increasing temperature under cw laser excitation of 2.33eV, which is higher than the band gap of GaSe slab. As shown in Fig. 2a, the PL spectrum shifts to lower energy as the temperature increases from 10K to 40K. Around 50K, a higher energy peak around 2.102eV labeled as "A" emerges, which is absent below 50K. As temperature goes up to 290K, the peak A shifts to lower energy, and then merges together with other PL peaks. At room temperature, only one PL peak around 2.00eV can be observed. We fit the PL spectrum under 60K with 4 peaks, and label the two highest energy peaks as "A" and "B". It is shown in Fig. 2b that the energy difference between peaks A and B is around 39meV at 60K, which is comparable with the energy difference (34meV) between high energy PL peak (2.086eV) and exciton absorption peak (2.12eV) at 10K shown in Fig.1a. Besides, peak A has a FWHM of 10meV, seven times smaller than that of peak B. Therefore, we could attribute this peak A (around 2.102eV) emerging at 60K to the free exciton and peak B at lower energy to the bound exciton. The temperature dependence of this free exciton emission is shown in Fig. 2c, where red spots represent PL peak position of free exciton with FWHMs labeled by red bar, whereas black squares represent exciton absorption peak with FWHMs labeled by black bar at various temperatures. Peak position of this free exciton emission changes from 2.10eV with a FWHM of 80meV at 60K to 2.0eV with a FWHM of 47meV at 290K. And exciton peak of absorption spectrum changes from 2.11eV with a FWHM of 9meV at 70K to 2.01eV with a FWHM of 14meW at 290K. Two peaks almost overlap in the temperature ranging from 50K to 290K. Both of them exhibit the same temperature dependence, which can be fitted by Varshni empirical relationships. All evidences tend to demonstrate that the peak A of PL spectrum stems from radiative recombination of free excitons. Because the probability of thermal dissociation effect in bound excitons increases with the increased temperature, more and more free excitons are released from the bound states. Therefore, the PL intensity of free excitons becomes relatively higher than that of bound excitons as temperature rises (Fig. 2a). So under room temperature, most of PL comes from the radiative recombination of free exciton. Below 50K, photoexcited free excitons are quickly captured by defect/impurity centers to form bound states, and consequently radiative recombination of free excitons is difficult to be detected. These results provide another potential cause for energy difference in PL and absorption spectrum at low temperature.

We also investigated temperature-dependent integrated PL intensity under the same experimental conditions, including excitation power as well as exposure time. As shown in Fig. 3a, the PL intensity dramatically drops by 2 orders of magnitude as temperature rises from 10K to 50K, and then has a flat plateau above 60K. The temperature dependence below 50K indicates that some non-radiative processes dominate the PL intensity, which is activated by increasing temperature. In such a situation, PL intensity is proportional to the scattering rate of non-radiative process- $I \propto \exp(-E/k_B T)$ where E is the activated energy. As shown in Fig.3b, the black points are integrated intensity at different temperature, and the red curve is a fitting result based on the above equation, from which we extract E=0.5meV. Because of the layered structure of GaSe slab, stacking faults and dislocation ofter occur in the samples. Therefore, these defect states form the non-radiative center, which induce the quenching of PL intensity as temperature rises.

Fig. 4a shows lifetime of the main PL peak at different temperatures. All PL lifetime data were collected around PL peak with bandwidth of 10nm. Due to limitation of the repetition frequency of pulse excited light, the maximum span of PL lifetime is around 12.5ns. The black curve in Fig.4a shows the PL lifetime as a function of delay time at 10K. Non-zero PL intensity (around 6000 counts) before time zero indicates that PL lifetime at 10K is longer than 12.5ns, which is the limitation of our setup. By considering non-zero PL intensity before time zero and fitting PL data by bi-exponential function at 10K, we deduce two time constants of 1.8ns and 55ns as shown in red curve of Fig.4a. The deduced fast process with time constant of 1.8ns can be attributed to bound exciton formation, which corresponds to the time that free excitons are captured by defects/impurities. And the slow process of 55ns is attributed to PL decay of bound exciton. At 260K, PL decay of free exciton only contains one component with time constant of 2.8ns. When temperature changes from 10K to 290K, PL lifetime (the slow process) sharply decreases from 55ns at 10K to 12ns at 80K, then PL lifetime approach a constant of 2.3ns above 80K. The trend of PL lifetime shown in Fig.4b is consistent with that of PL intensity, which further indicates that non-radiative process dominates at lifted temperature.

In summary, we observe a big energy difference between exciton absorption peak and PL peak below 50K. By analyzing the temperature dependence of PL spectra, we propose that the main PL process origin from emission of bound exciton at low temperature and the absorption peak results from absorption of free exciton at direct gap. By comparing PL intensity and lifetime under various temperatures, the active energy of non-radiative process is extracted to be around 0.5meV.

Acknowledgment. We thank Prof. Xiao-Dong Cui and Chun-Lei Yang for helpful discussions. This work is supported by National Natural Science Foundation of China (11204184).

1. Cao, T., Z. Li, and S.G. Louie, *Tunable Magnetism and Half-Metallicity in Hole-Doped Monolayer GaSe.* Physical Review Letters, 2015. **114**(23): p. 236602.
2. Tahara, H., Y. Ogawa, and F. Minami, *Non-Markovian Dynamics of Spectral Narrowing for Excitons in the Layered Semiconductor GaSe Observed Using Optical Four-Wave Mixing Spectroscopy.* Physical Review Letters, 2011. **107**(3): p. 037402.
3. Zhou, X., et al., *Strong Second-Harmonic Generation in Atomic Layered GaSe.* Journal of the American Chemical Society, 2015. **137**(25): p. 7994-7997.
4. Karvonen, L., et al., *Investigation of Second- and Third-Harmonic Generation in Few-Layer Gallium Selenide by Multiphoton Microscopy.* Scientific Reports, 2015. **5**: p. 10334.
5. Jie, W., et al., *Layer-Dependent Nonlinear Optical Properties and Stability of Non-Centrosymmetric Modification in Few-Layer GaSe Sheets.* Angewandte Chemie International Edition, 2015. **54**(4): p. 1185-1189.
6. Tang, Y., et al., *Optical and spin polarization dynamics in GaSe nanoslabs.* Physical Review B, 2015. **91**(19): p. 195429.
7. Minami, F., Y. Oka, and T. Kushida, *Effects of External Magnetic Fields on Optical Spin Orientation in GaSe.* Journal of the Physical Society of Japan, 1976. **41**(1): p. 100-108.
8. Halpern, J. in *Proceedings of the International Conference on Semiconductors*. 1966. Kyoto.
9. K. Aoyagi, A.M., G. Kuwabara, Y. Nishina, S. Kurita, T. Fukuroi, O. Akimoto, H. Hasegawa, M. Shinada and S. Sugano. in *Proceedings of the International Conference on Semiconductors*. 1966. Kyoto.
10. Baldereschi, A. and F. Bassani, *Landau Levels and Magneto-Optic Effects at Saddle Points.* Physical Review Letters, 1967. **19**(2): p. 66-68.
11. Brebner, J.L., *Magnetooptical Properties of GaSe near the Band Gap.* Canadian Journal of Physics, 1973. **51**(5): p. 497-504.
12. Watanabe, K., K. Uchida, and N. Miura, *Magneto-optical effects observed for GaSe in megagauss magnetic fields.* Physical Review B, 2003. **68**(15): p. 155312.
13. Mooser, E. and M. Schlüter, *The band-gap excitons in gallium selenide.* Il Nuovo Cimento B (1971-1996), 1973. **18**(1): p. 164-208.

14. Nüsse, S., et al., *Carrier cooling and exciton formation in GaSe.* Physical Review B, 1997. **56**(8): p. 4578-4583.
15. Dey, P., et al., *Biexciton formation and exciton coherent coupling in layered GaSe.* The Journal of Chemical Physics, 2015. **142**(21): p. 212422.
16. P. Fielding, G.F.a.E.M., Journ. Phys. Chem. Sol., 1959. **434**(8).
17. Mak, K.F., et al., *Atomically Thin ${\mathrm{MoS}}_{2}$: A New Direct-Gap Semiconductor.* Physical Review Letters, 2010. **105**(13): p. 136805.
18. Li, X., et al., *Controlled Vapor Phase Growth of Single Crystalline, Two-Dimensional GaSe Crystals with High Photoresponse.* Scientific Reports, 2014. **4**: p. 5497.

Figure 1 (a) Absorption and PL spectrum of GaSe slab at 10K. (b) Exciton energy peak position (black squares) as a function of temperature. The solid red line represents the fitting curve by using Varshni empirical relationship $E_g(T) = E_g(0) - \frac{AT^2}{T+B}$. (c) Muli-peak fitting of PL spectrum at 10K, and energy of these three peaks are identified as 2.086eV, 2.064eV and 2.032eV, respectively. (d) PL spectrum as a function of pump intensity ranging from 30uw to 600uw at 10K.

Figure 2 (a) Normalized PL spectra under different sample temperature ranging from 10K to 290K. (b) PL spectrum and muli-peak fitting curves at T=60K. (c) Absorption peak (black squares) and free exciton PL peak (red circles) as a function of temperature, where bars represent the FWHMs.

Figure 3 (a) PL spectra at various temperatures under the same experimental conditions, including excitation power, CCD optical integral time. (b) The integrated PL intensity as a function of temperature (black squares) and fitting curve (solid red curve) by using the equation $I \propto \exp(-E/k_B T)$

Figure 4 The PL lifetimes as a function of delay time and fitting curve at 8.5K (a) and at 260K (b). (c) PL lifetime under different sample temperature. (d) The PL lifetime (slow process), extracted from PL intensity vs delay time curve, as a function of temperature.

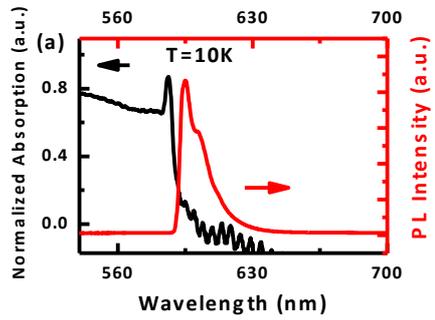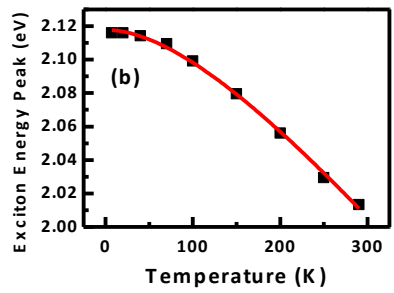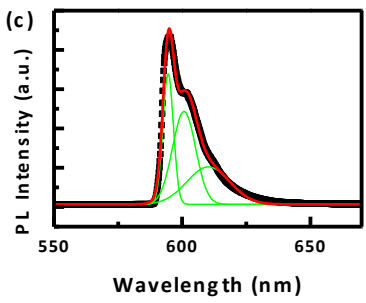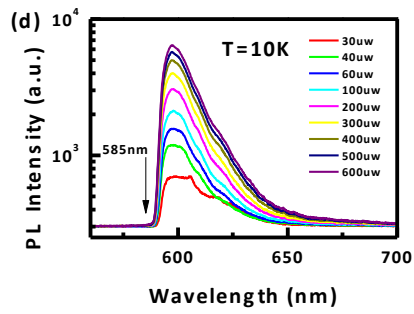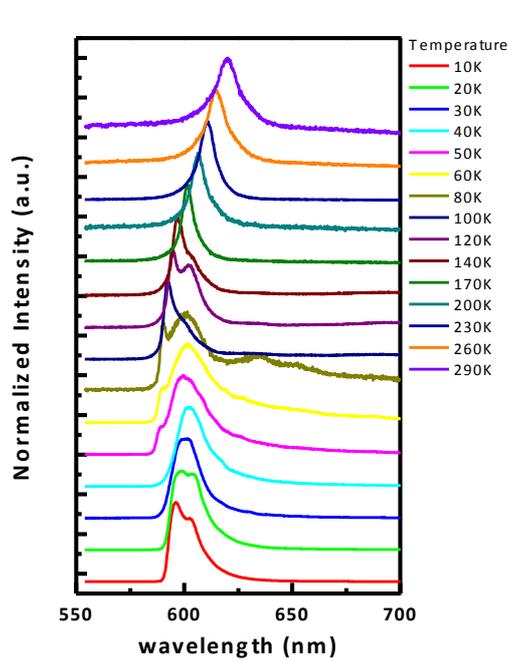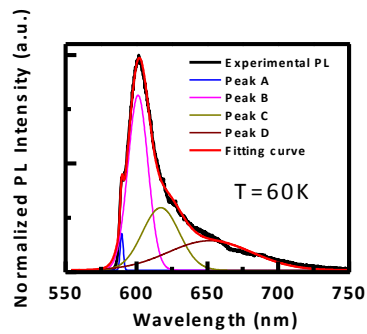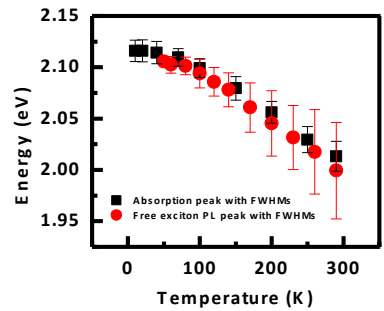

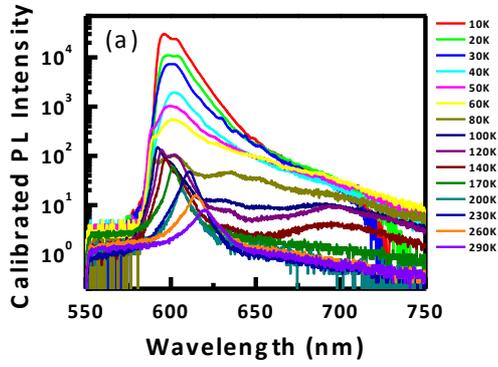
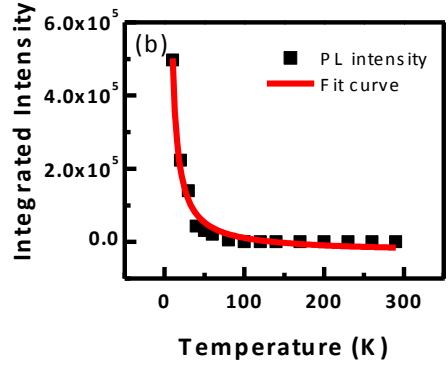

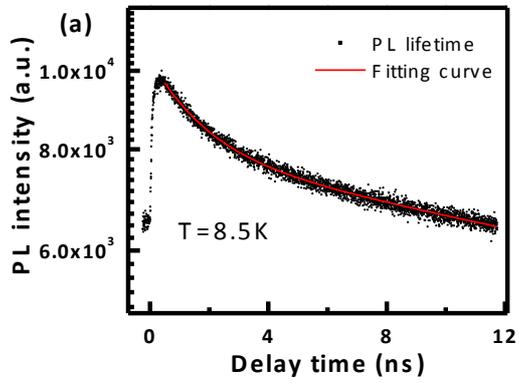
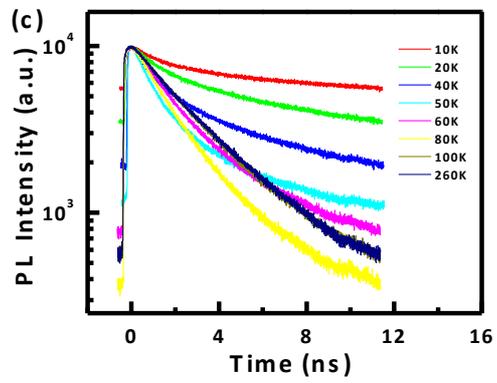

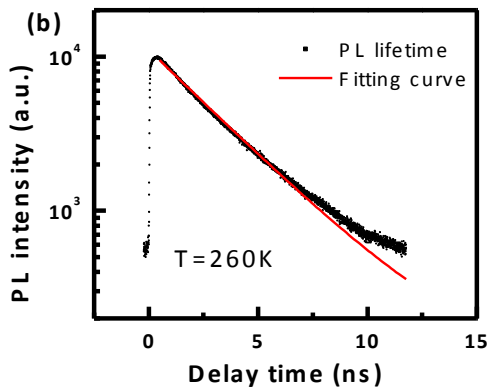
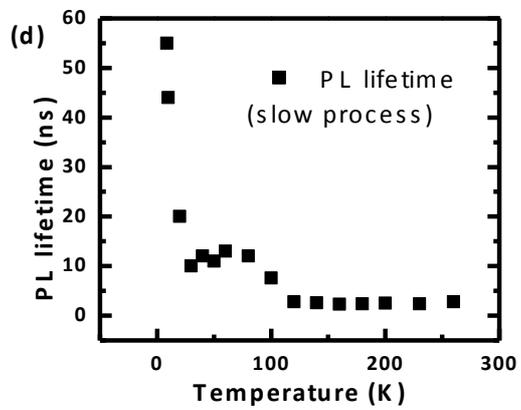